\documentclass[aip,amsmath,amssymb,graphicx,reprint]{revtex4-1}

\usepackage[version=3]{mhchem} 
\usepackage{multirow}
\usepackage[table]{xcolor}
\usepackage{graphicx} 

\begin{document}

\author{Paolo Bombelli}
\altaffiliation{These authors contributed equally to this work.}

\affiliation{Department of Biochemistry, University of Cambridge, Tennis Court Road, Cambridge CB2 1QW, United Kingdom}

\author{Thomas M\"uller}
\altaffiliation{These authors contributed equally to this work.}

\author{Therese W. Herling}
\affiliation{Department of Chemistry, University of Cambridge, Lensfield Road, Cambridge CB2 1EW, United Kingdom}

\author{Christopher J. Howe}
\email{ch26@cam.ac.uk}
\affiliation{Department of Biochemistry, University of Cambridge, Tennis Court Road, Cambridge CB2 1QW, United Kingdom}

\author{Tuomas P. J. Knowles}
\email{tpjk2@cam.ac.uk}
\affiliation{Department of Chemistry, University of Cambridge, Lensfield Road, Cambridge CB2 1EW, United Kingdom}

\title 
{A High Power-Density Mediator-Free Microfluidic Biophotovoltaic Device for Cyanobacterial Cells}

\begin{abstract}
  Biophotovoltaics has emerged as a promising technology for generating renewable energy since it relies on living organisms as inexpensive, self-repairing and readily available catalysts to produce electricity from an abundant resource - sunlight. The efficiency of biophotovoltaic cells, however, has remained significantly lower than that achievable through synthetic materials. Here, we devise a platform to harness the large power densities afforded by miniaturised geometries. To this effect, we have developed a soft-lithography approach for the fabrication of microfluidic biophotovoltaic devices that do not require membranes or mediators. \textit{Synechocystis sp.}~PCC 6803 cells were injected and allowed to settle on the anode, permitting the physical proximity between cells and electrode required for mediator-free operation. We demonstrate power densities of above 100~mW/m$^2$ for a chlorophyll concentration of $100~\mu$M under white light, a high value for biophotovoltaic devices without extrinsic supply of additional energy. 
\end{abstract}

\maketitle


Fuelling the ever-growing need for energy\cite{EIA2013} by fossil combustibles is expected to have dramatic, global consequences on climate and ecosystems. These environmental effects, in combination with the depletion of fossil fuel reserves, have led to a pressing need for developing technologies for harnessing renewable energy.\cite{Lewis2006,New2011} In this scenario, bio-electrochemical systems - such as microbial fuel cells\cite{Rabaey2005,Logan2006,Yang2011,Jiang2013} (MFCs) and biological photovoltaic cells\cite{Tsujimura2001,Rosenbaum2005,Pisciotta2010,Bombelli2011,Samsonoff2014} (BPVs) - may help to alleviate the present concerns by utilising living organisms as inexpensive, readily available catalysts to generate electricity. A particularly advantageous feature of BPVs is that they consist of living photosynthetic material that allows for continuous repair of photo-damage to key proteins.

Whereas MFCs use heterotrophic bacteria to convert the chemical energy stored in organic matter, BPVs use photosynthetic organisms capable of harnessing solar energy. In MFCs operating with \textit{Geobacter sulfurreducens}, the oxidation of acetate can proceed with a Coulombic efficiency of $\sim100\%$.\cite{Nevin2008} Nevertheless, the availability of acetate and other organic substrates is not endless which imposes a limiting factor to this approach. By contrast, in BPV-type systems, the conversion efficiencies of light into charges remain low ($\sim0.1\%$),\cite{McCormick2011} but the primary fuel (i.e., solar light) is virtually unlimited. Consequently, a significant research effort is required towards understanding which processes limit the performance of biophotovoltaic cells, both in terms of biophysics and engineering.

In this context, miniaturisation of BPVs provides highly attractive possibilities for high-throughput studies of small cell cultures, down to individual cells, in order to learn about differences in genetically identical organisms as well as to direct the evolution of efficient cell lines in bulk\cite{Carter2006,Bershtein2008,Keasling2008} and in microfluidics.\cite{Agresti2010} Furthermore, the distances which the charge carriers have to migrate within the devices can be shortened dramatically, reducing resistive losses in the electrolyte.\cite{Rabaey2005} The readily achievable conditions for laminar flow and sessile state of the anodophilic photosynthetic cells also permit operation without the use of a proton-exchange membrane.\cite{Choban2004,Kjeang2009,Wang2013,Ye2013}

To date, efforts have focussed on miniaturised microbial fuel cells.\cite{Chiao2006,Crittenden2006,Siu2008,Hou2009,Qian2009,Qian2011,Wang2011c,Hou2012,Ye2013,Jiang2013} In order to exploit the high power densities available through the decrease of the length scales of the charge transport and the decrease of the electrolyte volume, we have developed a simple fabrication method for microfluidic biophotovoltaic ($\mu$BPV) devices\cite{Chiao2006} that do not require an electron mediator or a proton-exchange membrane. Besides increasing efficiency and simplicity of the device, relinquishing mediator and membrane also reduces the cost of potential large-scale applications.\cite{Bond2003,Reguera2006,Malik2009,McCormick2011}

\begin{figure*}
  \includegraphics{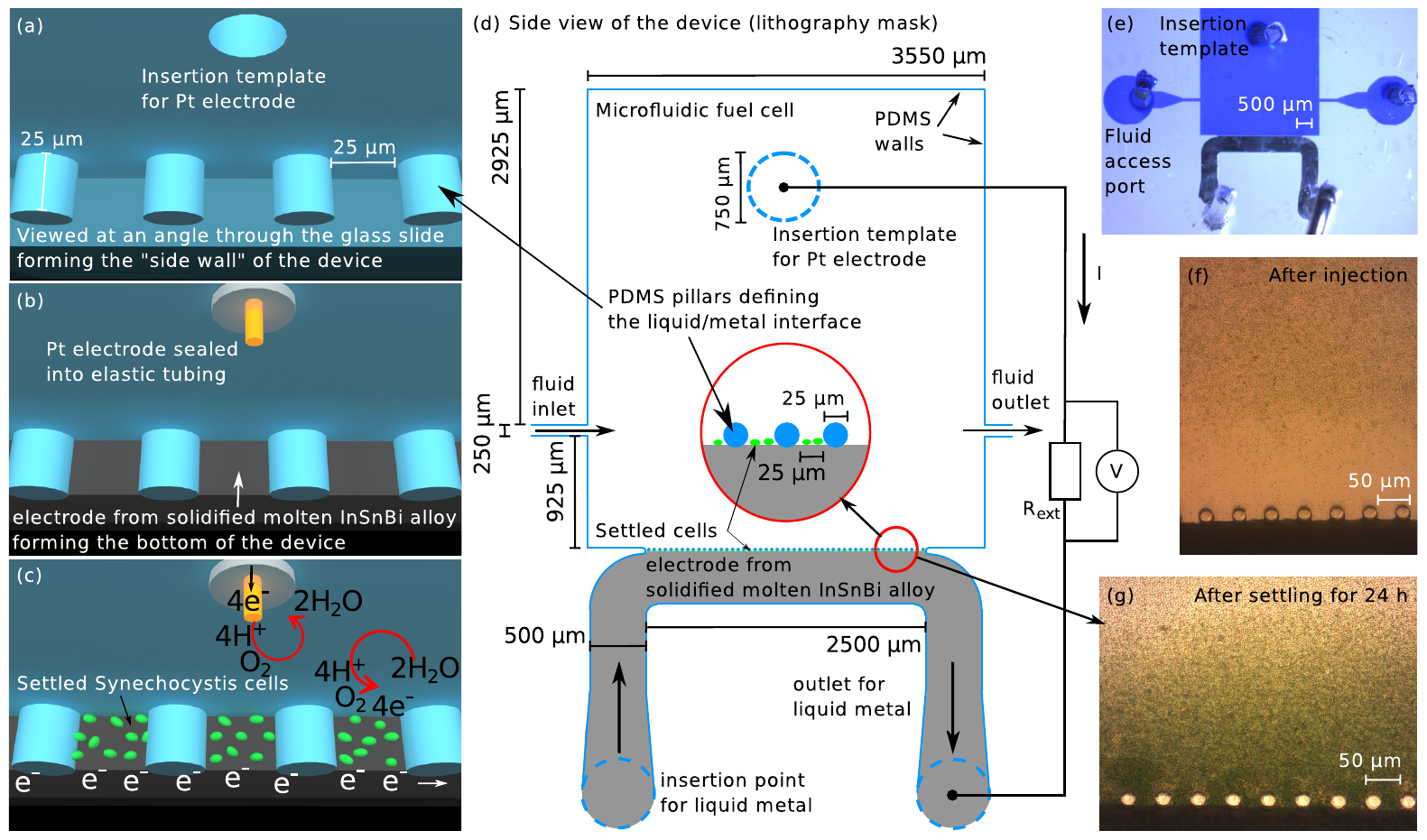}
  \caption{(a) Schematic of the device before insertion of the electrodes, seen at an angle through the glass slide. The lithographically defined PDMS pillars retain molten metal due to its surface tension, and the hole provides an opening for insertion of the \ce{Pt} electrode. (b) Model of the full device including platinum cathode and \ce{InBiSn} anode. (c) Schematic representation of the microfluidic biophotovoltaic device in action. \textit{Synechocystis} cells settled by gravity on the \ce{InBiSn} electrode deliver electrons to the latter by oxidising water. On the platinum cathode oxygen and hydrogen ions are supplied with electrons and combine to water, which closes the circuit. (d) Top view of the device design. (e) True-colour image of a device filled with a solution containing Coomassie blue to visualise the $25~\mu\textrm{m}$ high channels. (f) True-colour image of a device immediately after injection of \textit{Synechocystis} cells at a chlorophyll concentration of around $100~\mu$M. (g) True-colour image of a device filled with \textit{Synechocystis} cells that were allowed to settle on the anode during 24~h.}
  \label{sch:device}
\end{figure*}

We use soft lithography\cite{McDonald2002} to form microscopic channels which we equip using microsolidics\cite{Siegel2007} with a self-aligned electrode from a low-melting point alloy\cite{So2011,Li2013b,Herling2013} (InSnBi) and a platinum electrode sealed inside microfluidic tubing. A scheme of such a device is shown in Fig.~\ref{sch:device}(a-c), and the specific design including the external measurement circuit is presented in Fig.~\ref{sch:device}(d). True-colour microscopy photographs of a device filled with Coomassie blue, with freshly injected \textit{Synechocystis} cell, as well as with cells that have settled on the anode during 24 hours are shown in Fig.~\ref{sch:device}(e), (f), and (g), respectively. The possibility of omitting the mediator arises from the physical proximity of the settled cells and the anode which forms the bottom of the device, as well as the choice of electrode materials. The latter ensures that \ce{H}$^+$ is preferably reduced at the cathode since platinum catalyses this reaction.

The inherently small size (below 400~nL) of our microfluidic approach permits studies of minute amounts of biological material. Moreover, our $\mu$BPV works without any additional energy supply, such as inert gas purging to keep the anodic chamber anoxic and/or oxygen gas purging in the cathodic chamber to facilitate the reformation of water,\cite{Yagishita1997,Torimura2001,Tsujimura2001} or a bias potential applied to polarise the electrodes and improve the electron flux between anode and cathode.\cite{Malik2009}

The use of soft lithography allows for fast in-house prototyping and for the utilisation of the range of techniques developed for integrated circuits. Despite the small volumes contained in microfluidic devices, such approaches can be scaled up by parallelisation,\cite{Hou2012,Romanowsky2012} and the surface-to-volume ratio can be designed to outperform macroscopic approaches significantly.\cite{Wang2011c}

\section*{Results}
The microfluidic BPV device described here operates as a microbial fuel cell with submicroliter volume, generating electrical power by harnessing the photosynthetic and metabolic activity of biological material. Its anodic half-cell consists of sessile \textit{Synechocystis} cells - performing water photolysis (\ce{2H_2O}$\rightarrow$\ce{4H^+ +4e^- +O_2}) and subsequent ``dark'' metabolism - as well as an anode made from an InSnBi alloy and a light source.

\subsection*{Current and power analyses}
A $\mu$BPV was loaded with wild type \textit{Synechocystis sp.}~PCC 6803 cells (subsequently referred to as \textit{Synechocystis}) suspended in BG11 medium - supplemented with NaCl - at a final chlorophyll concentration of 100~nmol~Chl\,mL$^{-1}$. The exoelectrogenic activity of three biological replicates of sessile cells was characterised under controlled temperature conditions sequentially in the same device.

\begin{figure*}
  \includegraphics{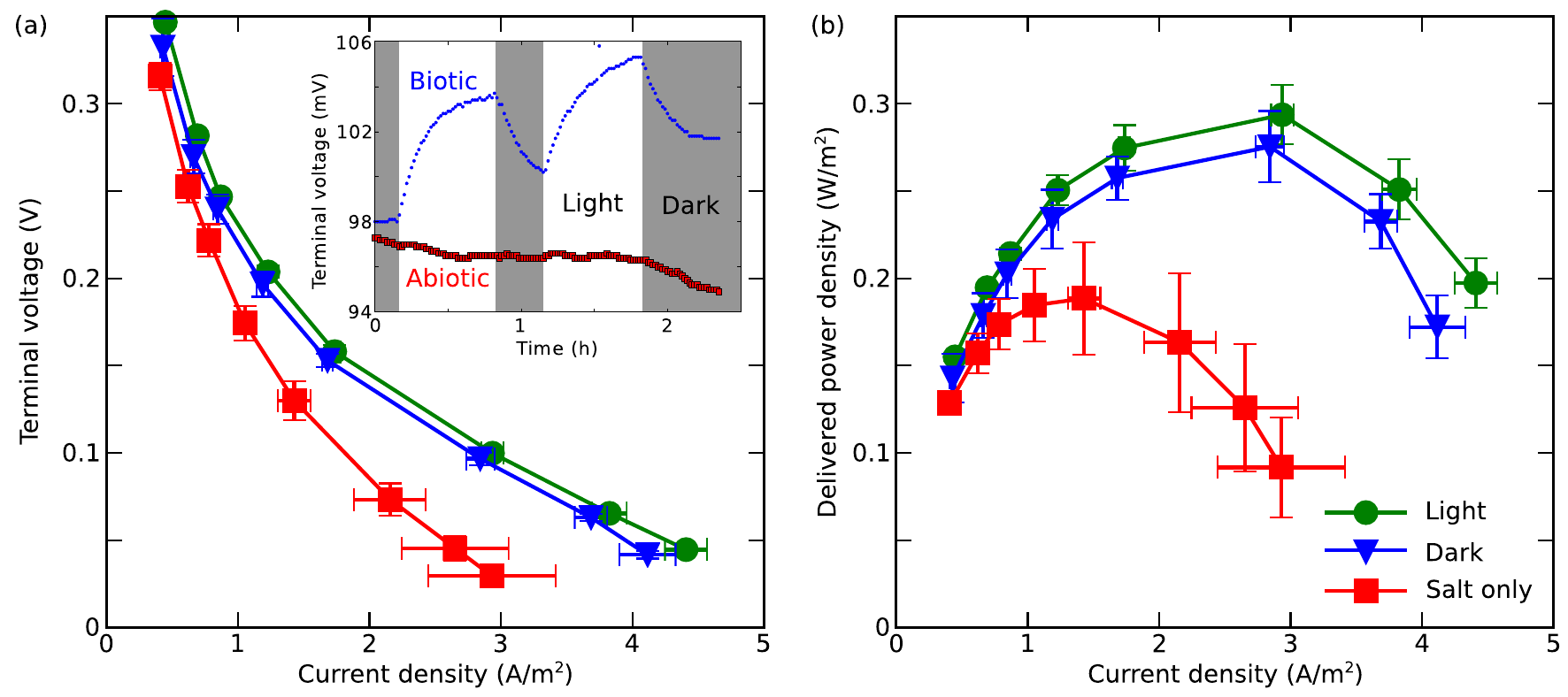} 
  \caption{(a) Comparison of the voltage output from the same microfluidic device loaded with salt medium only (BG11) or Syenechocystis cells in medium in the dark and with light. The $x$-axis has been converted to a current density through division of the measured current by the surface of the \ce{InSnBi} anode, and the error bars show the standard deviations for three consecutive, independent repeats on the same device. Inset: Response of the biophotovoltaic device as well as of an abiotic control under sequential illumination. (b) Power density generated by the microfluidic devices filled with salt or cells in dark/illuminated environment.}
  \label{gr:BPV}
\end{figure*}

The $\mu$BPV was rested for 24~hours, permitting the formation of cellular films on the anodic surface and stabilising the open circuit potential. Polarisation and power curves were then recorded by connecting different resistance loads to the external circuit in the dark or under illumination with white LED light (see Methods), and are shown in Fig.~\ref{gr:BPV}.

In the dark, significant power output was observed relative to the control sample containing no cells. This observation is consistent with the breakdown of stored carbon intermediates accumulated during the light period.\cite{Bombelli2011} The peak power output of $275\pm20~\textrm{mW}\,\textrm{m}^{-2}$ was established at a current density of $2840\pm110~\textrm{mA}\,\textrm{m}^{-2}$. Under illumination the microfluidic BPV loaded with \textit{Synechocystis} showed an increase in both current and power output. The peak power density was $P/A=294\pm17~\textrm{mW}\,\textrm{m}^{-2}$ established at a current of $2940\pm85~\textrm{mA}\,\textrm{m}^{-2}$. Crucially, both the dark and the light electrical outputs were significantly higher than the abiotic peak power output in this device of $189\pm32~\textrm{mW}\,\textrm{m}^{-2}$ established at a current of $1430\pm120~\textrm{mA}\,\textrm{m}^{-2}$, demonstrating that the power output from our devices originates from the biological activity of the cyanobacteria. 

From the linear slope at the high current side of the polarization curve as well as the from the external resistance for which maximal power transfer occurs we can estimate the internal resistance of the device to be around $2.2~\textrm{M}\Omega$ for the biotically loaded device and $1.4~\textrm{M}\Omega$ for the abiotic control (for further details see Supplementary Material).

The electrical output recorded from the abiotic control - possibly due to medium salinity\cite{Logan2006,Logan2009} and anodic oxidation - is taken into account when the power densities of biotic experiments are quoted. Specifically, subtracting the abiotic background yields a biotic output power density of $105~\textrm{mW}\,\textrm{m}^{-2}$. This number is halved when comparing to the full cross-sectional area of the device (including the inaccessible parts of the anode), and the power available per footprint area is ca.~$50~\mu\textrm{W}\,\textrm{m}^{-2}$.

\subsection*{Light response}
To demonstrate the photo-activity of the Synechocystis cells, the variation of the anode-cathode voltage as a response to repeated light stimulation was recorded over time (see inset of Fig.~\ref{gr:BPV}(a)). The external resistor was fixed at 100~M$\Omega$, and the voltage was sampled once per minute. Illumination by white LED light at $200~\mu\textrm{mol}\,\textrm{m}^{-2}\textrm{s}^{-1}$ resulted in a reproducible voltage increase at a rate of $21.7\pm4.7~\textrm{mV}\,\textrm{h}^{-1}$ with $\Delta\textrm{V}_\textrm{light-dark}=5.2\pm0.6~\textrm{mV}$. The time until the electrical outputs were stabilised was around one hour. We find that the baseline voltage levels change after illumination - most certainly due to a buildup and breakdown of intracellular metabolites.

From the measured spectrum of the light source (see Supplementary Information) we can determine the average wave number which corresponds to a wavelength of 570~nm. Thus the photon flux can be converted to an incident light intensity of $42~\textrm{W}\,\textrm{m}^{-2}$. Using these values we can extract a rough estimate for the efficiency of our BPV (energy output versus energy input) of around 0.25\% which compares favourably to previously achieved values.\cite{Chiao2006,McCormick2011,Lan2013} Note that light scattering on the glass surface and losses from the non perpendicular illumination angle would increase this number and hence it can be understood as a lower bound.

With such an illumination cycle, the light-driven electrical response of a device can be directly compared to dark conditions, proving the functionality of our $\mu$BPV. In addition, the abiotic control shows no variations in anode-cathode potential under similar illumination.

The difference between the power outputs under dark and illuminated conditions is consistent with previous studies of \textit{Synechocystis sp.}~PCC 6803.\cite{McCormick2011} Nevertheless, a direct comparison of the power output reported by McCormick \textit{et al.}~of around $0.12~\textrm{mW}\,\textrm{m}^{-2}$ with the peak value in excess of $100~\textrm{mW}\,\textrm{m}^{-2}$ demonstrated here emphasises the great potential of microfluidic approaches compared to macroscopic devices.

\subsection*{Variability of the abiotic characterisation}

In order to characterise the variability of the electrical behaviour of our $\mu$BPV, two further, lithographically identical devices were studied with abiotic loading (i.e., without photosynthetic cells). These devices were injected with BG11 media (with 0.25 M NaCl), and the current and power outputs were characterised under controlled temperature conditions.

\begin{figure}
\centering
  \includegraphics{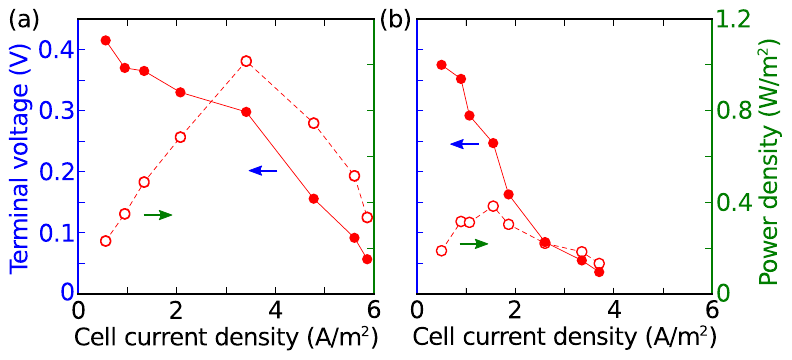}
  \caption{(a) and (b) Output voltage (filled circles, solid line, blue axis) and available power density (hollow circles, dashed line, green axis) as a function of current from two further abiotically loaded devices (BG11 cell medium supplemented with 0.25 M NaCl).}
  \label{gr:abiotic}
\end{figure}

Following 24~hours of stabilisation of the $\mu$BPV at open circuit potential, polarisation and power curves (see Fig.~\ref{gr:abiotic}) were generated by applying different resistance loads to the external circuit in the dark. In different devices, the abiotic peak power density outputs vary from around 0.2 to $1~\textrm{W/m}^2$ and were established at current densities of 1.5 and $3.5~\textrm{A/m}^2$, respectively. The large variation in device output between different devices stems from the variable position and shape of the cathode which is not lithographically defined in our current designs. Device improvements at this level may well provide a straightforward route to further improvement of the output power. Crucially, no major changes in current and power outputs were observed upon exposure to white light (see inset of Fig.~\ref{gr:BPV}(a)).

\subsection*{Comparison with recent literature}
The exceptionally high power density in excess of $100~\textrm{mW}\,\textrm{m}^{-2}$ after subtraction of the abiotic background has been facilitated by the physical proximity of the cells to the anode allowing for operation without a proton-exchange membrane, which in turn leads to a low internal resistance in the device, as well as by the microscopic size of the anodic chamber allowing for a large ratio of active surface to volume. In macroscopic bio-electrochemical systems by contrast, parameters such as mass transport, reaction kinetics and ohmic resistance are expected to have detrimental effect on the electrical output.\cite{Rabaey2005,Wang2011c} 

\begin{table*}[htb]\footnotesize
\begin{center}
\begin{tabular}{ l c c c c c c c}
 
   \multirow{2}{*}{Study} & \multicolumn{1}{c}{\textbf{$P_\textrm{out}$}} & \multicolumn{1}{c}{AAA} & \multicolumn{1}{c}{ACV} & Anode/ & \multirow{2}{*}{Mediator} & Photosynthetic\\
  & \textbf{mW/m$^2$} & \multicolumn{1}{c}{mm$^2$} & \multicolumn{1}{c}{$\mu$L} & Cathode & & organism\\
  \hline\noalign{\smallskip}
  \multirow{2}{*}{Chiao 2006\cite{Chiao2006}} & \textbf{\multirow{2}{*}{0.0004}} & \multirow{2}{*}{50} & \multirow{2}{*}{4.3} & Au/ & Methylene & \multirow{2}{*}{\textit{Anabaena sp.}}\\ & & & & N-Au - csc & blue & \\\noalign{\smallskip}
  \multirow{2}{*}{Bombelli 2011\cite{Bombelli2011}} & \textbf{\multirow{2}{*}{1.2}} & \multirow{2}{*}{80} & \multirow{2}{*}{150} & \multirow{2}{*}{ITO/N-CPt} & \multirow{2}{*}{\ce{K_3[Fe(CN)_6]}} & \textit{Synechocystis sp.} \\ & & & & & & PCC 6803\\\noalign{\smallskip}
  \multirow{2}{*}{McCormick 2011\cite{McCormick2011}} & \textbf{\multirow{2}{*}{10}} & \multirow{2}{*}{1'300} & \multirow{2}{*}{12'600} & ITO/ & \multirow{2}{*}{free} & \textit{Synechococcus sp.} \\ & & & & Pt-coated glass & & WH 5701\\
  \multirow{2}{*}{Thorne 2011\cite{Thorne2011}} & \textbf{\multirow{2}{*}{24}} & \multirow{2}{*}{230} & \multirow{2}{*}{2'300} & \multirow{2}{*}{FTO/Carbon cloth} & \multirow{2}{*}{\ce{K_3[Fe(CN)_6]}} & \multirow{2}{*}{\textit{Chlorella vulgaris}}\\ & & & & & &\\\noalign{\smallskip}
  Bombelli 2012\cite{Bombelli2012} & \textbf{0.02} & 2'000 & 20'000 & ITO/Pt-C & free & \textit{Oscillatoria limnetica} \\ \noalign{\smallskip}
  \multirow{2}{*}{Madiraju 2012\cite{Madiraju2012}} & \textbf{\multirow{2}{*}{0.3}} & \multirow{2}{*}{1'500} & \multirow{2}{*}{60'000} & \multirow{2}{*}{Carbon fibre} & \multirow{2}{*}{free} & \textit{Synechocystis sp.} \\ & & & & & & PCC 6803\\\noalign{\smallskip}
  Bradley 2013\cite{Bradley2013} & \textbf{0.2} & 1'300 & 31'500 & ITO/N-CPt & \ce{K_3[Fe(CN)_6]} & \textit{Synechocystis TM}\\\noalign{\smallskip}
  \multirow{2}{*}{Lan 2013\cite{Lan2013}} &\textbf{\multirow{2}{*}{13}}& \multirow{2}{*}{4'600} & \multirow{2}{*}{$5\times10^5$} & \multirow{2}{*}{Pre-treated graphite/csc} & \multirow{2}{*}{\ce{K_3[Fe(CN)_6]}} & \textit{Chlamydomonas} \\ & & & & & & \textit{reinhardtii} \\\noalign{\smallskip}
  Lin 2013\cite{Lin2013} & \textbf{10} & 2'100 & $10^6$ & Au mesh/Graphite cloth & free & \textit{Spirulina platensis}\\\noalign{\smallskip}
  \multirow{2}{*}{Luimstra 2013\cite{Luimstra2013}} & \textbf{\multirow{2}{*}{6}} & \multirow{2}{*}{1'400} & \multirow{2}{*}{70'000} & PPCP/ & \multirow{2}{*}{free} & \textit{Pauschulzia} \\ & & & & Carbon cloth with Pt &  & \textit{pseudovolvox} \\\noalign{\smallskip}
  \multirow{2}{*}{Sekar 2014\cite{Sekar2014}} & \textbf{\multirow{2}{*}{35}} & \multirow{2}{*}{2.5} & \multirow{2}{*}{n/a} & CNTCP/ & \multirow{2}{*}{free} & \multirow{2}{*}\textit{Nostoc sp.} \\ & & & & Laccase on CNTCP &  &  \\\noalign{\smallskip}
  \multirow{2}{*}{Sekar 2014\cite{Sekar2014}} & \textbf{\multirow{2}{*}{100}} & \multirow{2}{*}{2.5} & \multirow{2}{*}{n/a} & CNTCP/ & \multirow{2}{*}{BQ} & \multirow{2}{*}\textit{Nostoc sp.} \\ & & & & Laccase on CNTCP &  &  \\\noalign{\smallskip}
  \multirow{2}{*}{This study} & \textbf{\multirow{2}{*}{105}} & \multirow{2}{*}{0.03} & \multirow{2}{*}{0.4} & \multirow{2}{*}{InSnBi alloy/Pt} & \multirow{2}{*}{free} & \textit{Synechocystis sp.} \\ & & & & & & PCC 6803\\
  
\end{tabular}
\parbox{\textwidth}{\caption{\small List of biophotovoltaic devices from the recent literature - including previous microfluidic approaches - that do not require additional energy input. The abbrevations used are anodic active area (AAA), anodic chamber volume (ACV), Nafion film over the cathodic chamber and Au cathode (N-Au), chemical sacrificial cathode (csc), carbon-platinum cathode impregnated on one side with Nafion (N-CPt), carbon paper coated with a thin layer of platinum (Pt-C), indium tin oxide (ITO), fluorine doped tin oxide (FTO), carbon paint with polypyrrole (PPCP), carbon nanotubes on carbon paper (CNTCP), and benzoquinone (BQ). \textit{Synechocystis TM} refers to mutant strains of the cyanobacterium \textit{Synechocystis sp.}~PCC 6803 where the three respiratory terminal oxidase complexes had been inactivated.}\label{Tab:Values}}
\end{center}
\end{table*}

For a specific comparison, Tab.~\ref{Tab:Values} gives an overview of the power densities as well as technical specifications of intrinsic BPVs (i.e., requiring no external energy) characterised in the recent literature, including an instance with an additional enzymatic cathode.\cite{Sekar2014} While there are many aspects influencing the performance of a BPV, such as surface-to-volume ratio, photosynthetic organism, and electrode material, one can observe a trend that generally the mediator-free approaches surpass their counterparts that rely on electron mediators diffusing over large distances. It should be mentioned that many of the studies listed in Tab.~\ref{Tab:Values} were not intended to improve on output power. We also note that higher power densities have been observed\cite{Tsujimura2001} when extrinsic energy was supplied.

\section*{Discussion}
In summary, we have described a microfluidic design for a mediator-less, membrane-free bio-photovoltaic device. Electrical characterisation of devices loaded with \textit{Synechocystis sp.}~PCC 6803 revealed peak power densities in excess of $100~\textrm{mW/m}^2$. In spite of the low power available per footprint area (currently of the order of $50~\mu\textrm{W/m}^2$) the promising performance and the simple fabrication process demonstrate the potential of our approach for generating biological solar cells with microfluidics. 

Our approach is applicable to any photosynthetic organism forming biofilms. Furthermore, using the strategy presented in this work, further improvement of the power output should be readily achievable through reduction of the distance between anode and cathode and increase of the channel height. This flexibility in device geometry and the possibility of \textit{in-situ} electroplating of the anode underline the versatility of soft-lithography as a means for generating biophotovoltaic cells.

Options for enhanced miniaturisation open pathways for the study of small cell cultures containing as little as tens of cells for rapid screening of electrochemically active microbes in the context of directed evolution.

\begin{small}
\section*{Methods}

\subsection*{Device fabrication}

Devices were fabricated to a height of $25~\mu\textrm{m}$ using standard soft lithography\cite{McDonald2002} for polydimethylsiloxane (PDMS) on glass. The designs include an array of $25~\mu\textrm{m}$ wide PDMS pillars spaced by $25~\mu\textrm{m}$ in order to allow for insertion of molten solder\cite{So2011,Li2013b} (Indalloy 19, Indium Corporation, Clinton NY, USA) on a hotplate set to $79~^\circ\textrm{C}$. Solidification of this \ce{InBiSn} alloy upon removal from the heat yields self-aligned wall electrodes using a single lithography step.\cite{Herling2013} This process is illustrated in Fig.~\ref{sch:device}(a) and (b). The cathode is constructed by inserting a strip of platinum wire of $100~\mu\textrm{m}$ diameter through polyethylene tubing (Smiths Medical; 800/100/120; the same as used for contacting microfluidic devices in general) and sealing off both ends of the tubing with epoxy glue. Inserting this tube through a previously punched hole in the device generates a sealed electrical connection and is indicated by the orange wire (Pt) inside a white cylinder (tubing) in the scheme in Fig.~\ref{sch:device}(b). Note that this method for electrode fabrication also allows for straightforward exchange of the cathode material, which would be beneficial for \textit{in-situ} electroplating the \ce{InBiSn} alloy.

During settling and operation, the BPVs are oriented such that the bottom of the device is formed by the anode, and the glass slide as well as the pdms forming the side and top walls.

The total volume above the anode is below 400~nL, significantly reducing the consumption of biological material and chemicals of each experiment compared to macroscopic approaches.

\subsection*{Electrode Area}
The accessible surfaces of these electrodes are ca.~$A\sim 2.5~\textrm{mm}/2\times 25~\mu\textrm{m}\approx 0.03~\textrm{mm}^2$ for the anode (only approximately one half of the total metal area is accessible due to the PDMS pillars) and of the order of $0.6~\textrm{mm}^2$ for the cathode, assuming the available length of the Pt wire to be 2~mm. Note that the majority of the cathode lies inside the cavity of the insertion template. If one were to consider the entire horizontal cross-section of the device, the according area would double to $0.06~\textrm{mm}^2$, and the footprint of the device is at present around $60~\textrm{mm}^2$ including the access ports for fluid injection. This latter number can be reduced straightforwardly by more than one order of magnitude by redesigning the inlet ports. 

\subsection*{Cell culture and growth}
A wild-type strain of \textit{Synechocystis sp.}~PCC 6803 was cultivated from a laboratory stock.\cite{Bombelli2011} Cultures were grown and then analysed in BG11 medium\cite{Rippka1979} supplemented with 0.25~M NaCl. All cultures were supplemented with 5~mM NaHCO$_3$ and maintained at $22\pm2~^\circ\textrm{C}$ under continuous low light (ca.~$50~\mu\textrm{mol}\,\textrm{m}^{-2}\textrm{s}^{-1}$) in sterile conditions. Strains were periodically streaked onto plates containing agar ($0.5-1.0\%$) and BG11 including NaCl, which were then used to inoculate fresh liquid cultures. Culture growth and density were monitored by spectrophotometric determination of chlorophyll content. Chlorophyll was extracted in $99.8\%$ (v/v) methanol (Sigma-Aldrich, Gillingham, UK) as described previously.\cite{Porra1989}

\subsection*{Cell injection and settling}
First, the devices were filled with culture medium (BG11 with 0.25~M NaCl) and any air bubbles were removed by means of syringes attached via elastic polyethylene tubing (Smiths Medical; 800/100/120). \textit{Synechocystis} cells suspended in BG11 (supplemented with NaCl) were then injected at a concentration of $100~\mu\textrm{M}$ chlorophyll. Maintaining the devices for 24~h at an orientation in which the metal alloy anode forms the bottom allows the cells to sediment on the electrode by gravity. This process creates a closely-spaced interface allowing the electrons to be transmitted to the anode (see Fig.~\ref{sch:device}(c) and (g)) and thus favouring mediator-free operation. Throughout all experiments, the syringes are kept attached in order to prevent drying out of the BPV.

The complete device design used for the photolithography mask is presented in Fig.~\ref{sch:device}(d), and a microscopy photograph of a device coloured with Coomassie blue is shown in Fig.~\ref{sch:device}(e). Furthermore, a picture of an array of devices is provided in the supplementary material.

\subsection*{Microfluidic BPV measurement and illumination}
In principle, the optimal way of extracting the voltage output of our biophotovoltaic device would be to determine the half-cell potentials individually by integrating reference electrodes into the devices. Since this is challenging in microfluidic devices,\cite{Shinwari2010} we have instead measured the terminal voltage of our BPV which does not offer insight into the potentials of the complex half-cell reactions but provides an accurate measurement for the power delivered to an external load.

Polarisation curves were acquired by recording the terminal voltage $V$ under pseudo steady-state conditions\cite{Logan2006} with variable external loads ($R_\textrm{ext}$) and plotting the cell voltage as a function of current density (current per unit anodic area). Typically, a time span of around 20~min was sufficient for a stable output (see Supplementary Fig.~2). The resistance values ranged from 24.8~M$\Omega$ to 324~k$\Omega$ (24.8, 13, 9.1, 5.3, 2.9, 1.1, 0.547, and 0.324~M$\Omega$), where the internal resistance of the digital voltmeter of  $100~\textrm{M}\Omega$ has been taken into account. Voltages were recorded using an UT-70 data logger (Uni-Trend Limited, Hong Kong, China). The current delivered to the load was calculated from Ohm's law
\begin{equation}\label{Eq:Ohm}
V=R_\textrm{ext}I,
\end{equation}
and the power $P$ is given by
\begin{equation}\label{Eq:Pow}
P=V^2/R_\textrm{ext}.
\end{equation}

Based on the polarisation curves, power curves were obtained for each system by plotting the power per unit area or power density $P/A$ as a function of current density. These power density curves were further used to determine the average maximum power output for the microfluidic BPV system and the negative control. For all measurements, alligator clamps and copper wire served as connections to anode and cathode, and the temperature was kept at $22\pm2~^\circ\textrm{C}$.

To characterise the light response, artificial light was provided by a warm white LED bulb (Golden Gadgets, LA2124-L-A3W-MR16), maintained at a constant output of $200~\mu\textrm{mol}\,\textrm{m}^{-2}\textrm{s}^{-1}$ at the location of the BPVs. A measured spectrum of the light source is shown in the supplementary material. Light levels were measured in $\mu\textrm{mol}\,\textrm{m}^{-2}\textrm{s}^{-1}$ with a SKP 200 Light Meter (Skye Instruments Ltd, Llandrindod Wells, UK).

The photo-active cells were illuminated through the glass slide forming the bottom of the device, resulting in an almost parallel angle of incidence on the cell layer. This geometry does lead to a decreased light intensity on the cells, which may be compensated for by using a more powerful light source in studies of photosynthetic materials or by altering the geometric arrangement of the devices when harnessing actual sunlight.

\end{small}

\section*{References}

\begin{small}
\section*{Acknowledgements}

We gratefully acknowledge financial support from the Biotechnology and Biological Sciences Research Council (BBSRC), the Engineering and Physical Sciences Research Council (EPSRC), the European Research Council (ERC), the EnAlgae consortium (http://www.enalgae.eu/), as well as the Swiss National Science Foundation (SNF).

\newpage
\section*{Supplementary Information}

\subsection{Device Handling}
Figure \ref{gr:SupPic} illustrates the operation of the device. After the fabrication of the device and the electrodes, the cell medium is injected using plastic syringes, and all air is removed by applying pressure on the fluid inlet and outlet. Thereafter, the elastic tubing is cut on one side and the cells are injected through another syringe. The syringes are then left attached to prevent drying of the device. Copper wires are soldered to the electrodes to provide electric connections. Finally, the device is positioned such that the anode forms the bottom and the cells sediment on it under the influence of gravity.

\begin{figure}[bp]
  \includegraphics[width=0.45\textwidth]{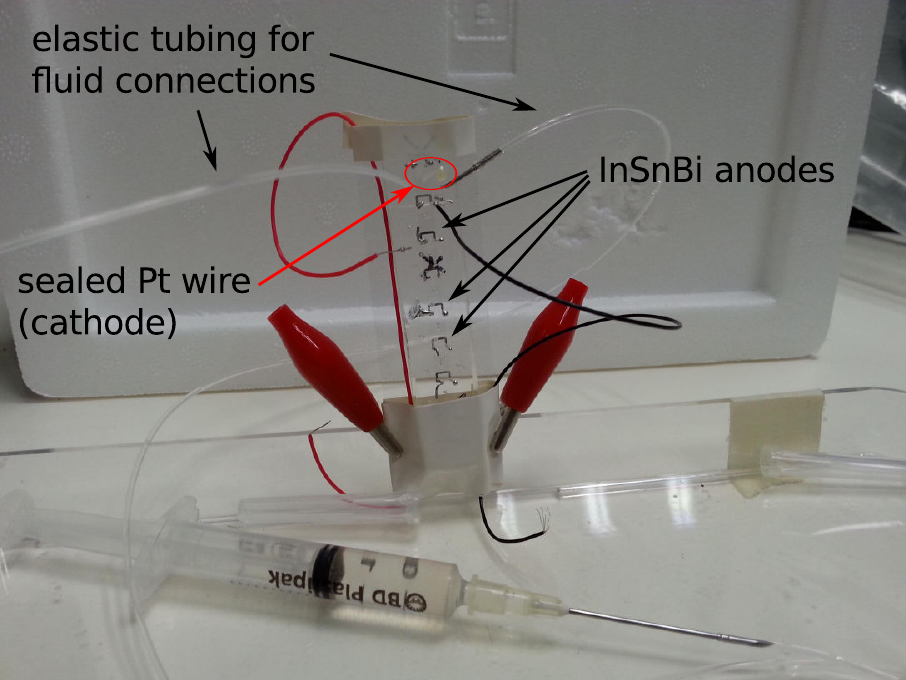} %
  \caption{Photograph of an array of devices. Anodes were fabricated for all cells, the sealed cathode and the fluid connectors are inserted into the topmost device.}
  \label{gr:SupPic}
\end{figure}

A magnified version of a device filled with a Coomassie blue solution is shown in Fig.~\ref{gr:DevPic}(a), and a true-colour microscopy image of \textit{Synechocystis sp.}~PC 6803 cells settled on the alloy anode is presented in Fig.~\ref{gr:SupPic}(b).

\begin{figure}
  \includegraphics[width=0.45\textwidth]{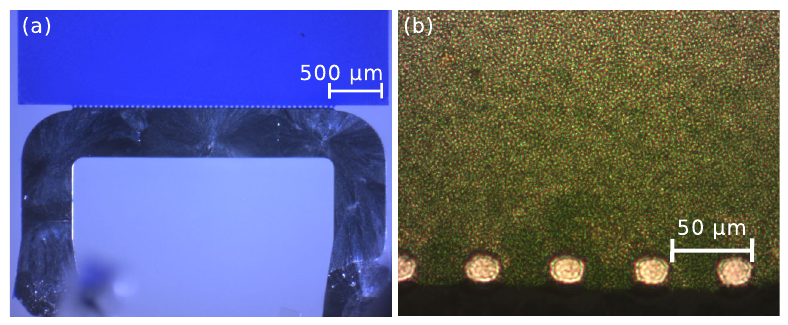} %
  \caption{(a) True-colour microscopy image of the indium alloy electrode and the fluid chamber filled with a Coomassie blue solution. (b) True-colour microscopy image of \textit{Synechocystis sp.}~PC 6803 cells settled on the electrode during 24~h.}
  \label{gr:DevPic}
\end{figure}

\subsection{Device stability}

In order to assess the stability of our devices, a sample was loaded with cell medium and its power output was measured during more than 25~hours (see Fig.~\ref{gr:control}). After a sharp decrease in the first minutes, the device was stable for a period in excess of 24~hours.
\begin{figure}
  \includegraphics[width=0.45\textwidth]{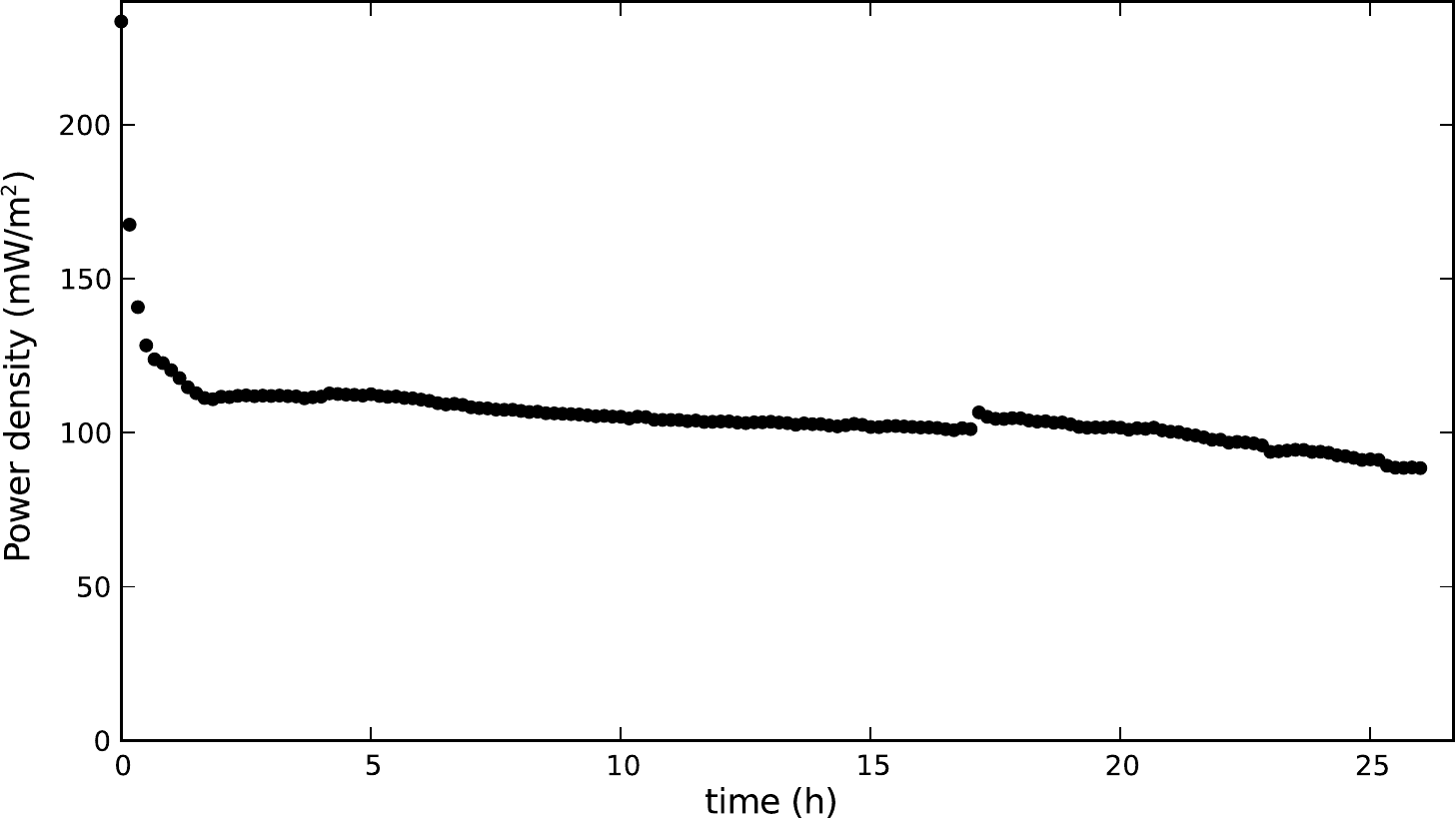} %
  \caption{Power output of an abiotically loaded device over an extended period of time, measured every 6~minutes.}
  \label{gr:control}
\end{figure}

\subsection{Anode Material}
The low-melting point solder Indalloy 19 (Indium Corporation, Clinton NY, USA) - composed of $51\%$ indium, $32.5\%$ bismuth, and $16.5\%$ tin - has been chosen due to its simplicity for generating self-aligned wall electrodes in microfluidic devices.\cite{So2011,Li2013,Herling2013} The melting point at $60~^\circ$C enables straightforward insertion into the device on a hot plate at $79~^\circ$C, with the liquid metal patterned by polydimethylsiloxane pillars due to its surface tension. Upon removing the devices from the hot plate the alloy solidifies forming a solid electrode. 

The standard potentials of each of the constituent metals are\cite{Bard1985}
\begin{align}
&\mathrm{In}  \leftrightarrow  \mathrm{In}^+\,+\,e^- \qquad & -0.13~\textrm{V},\\
&\mathrm{In}  \leftrightarrow  \mathrm{In}^{3+}\,+\,3e^- \qquad & -0.34~\textrm{V},\\
&\mathrm{Bi}  \leftrightarrow  \mathrm{Bi}^{3+}\,+\,3e^- \qquad & +0.32~\textrm{V},\\
&\mathrm{Sn}  \leftrightarrow  \mathrm{Sn}^{2+}\,+\,2e^- \qquad & -0.14~\textrm{V}.
\end{align}
These potentials are below the value for the oxidation of hydrogen to water at the cathode ($+1.23~\textrm{V}$), and therefore it is quite possible that oxides such as, for instance, \ce{In_2O_3}, \ce{Bi_2O_3}, \ce{Bi_2Sn_2O_7}, or \ce{SnO_2} are forming on the anode. We have subtracted this oxidative current from our power estimates and did not see any significant deterioration in performance in a control over a time span of 25~hours (Fig.~\ref{gr:control}).

\subsection{Estimates of the internal resistance}

Measuring the voltage drop over an external resistor attached to a source yields the terminal voltage which is smaller than the actual cell voltage due to the internal resistance of the source
\begin{equation}
V_\textrm{terminal}\equiv V = V_\textrm{cell}-R_\textrm{int}I=V_\textrm{cell}-\frac{R_\textrm{int}}{R_\textrm{ext}}V.
\end{equation}
Therefore,
\begin{equation}
V=\frac{V_\textrm{cell}}{1+R_\textrm{int}/R_\textrm{ext}}.
\end{equation}
Since the $IV$-characteristics are not linear, the internal resistance of the cell, or - more likely - its output voltage, depends on the current drawn. Nevertheless, from the linear part of the polarisation curve at high currents (Fig. 2(a) in the main text) we can estimate the internal resistance to amount to
\begin{equation}
\frac{\Delta V}{\Delta I}\approx \frac{158-45~\textrm{mV}}{(4.4-1.7~\textrm{A/m}^2)*0.03~\textrm{mm}^2}=1.4~\textrm{M}\Omega
\end{equation}
for the biotically loaded device and 
\begin{equation}
\frac{\Delta V}{\Delta I}\approx \frac{130-30~\textrm{mV}}{(2.9-1.4~\textrm{A/m}^2)*0.03~\textrm{mm}^2}=2.2~\textrm{M}\Omega
\end{equation}
for the abiotic control. Note that the internal resistance decreased by $1/3$ with the addition of the cyanobacteria. Furthermore, since maximal power transfer to the external load is observed when the load resistance is matched to the internal resistance of the cell, we can double-check the above values by comparison to Fig. 2(b). There, the maximum power is observed for external resistances of $1.1~\textrm{M}\Omega$ and $2.9~\textrm{M}\Omega$ for biotic and abiotic filling, respectively. These values are in close agreement with the estimates from the polarisation curves.

\subsection{Light source}

In Fig.~\ref{gr:Spec} we present the measured spectrum of the lamp we have used to illuminate our biophotovoltaic cells. From these data, we can also extract the weighted average wave number to be $10^7~\textrm{m}^{-1}$ which corresponds to a wavelength of around 570~nm. Therefore, the average energy per photon is $3.5\times10^{-19}~\mathrm{J}$, and the measured photon flux of $200~\mu\textrm{mol}/\textrm{m}^2/\textrm{s}$ yields an illumination intensity at the location of the devices of $42~\textrm{W/m}^2$.
\begin{figure}
  \includegraphics{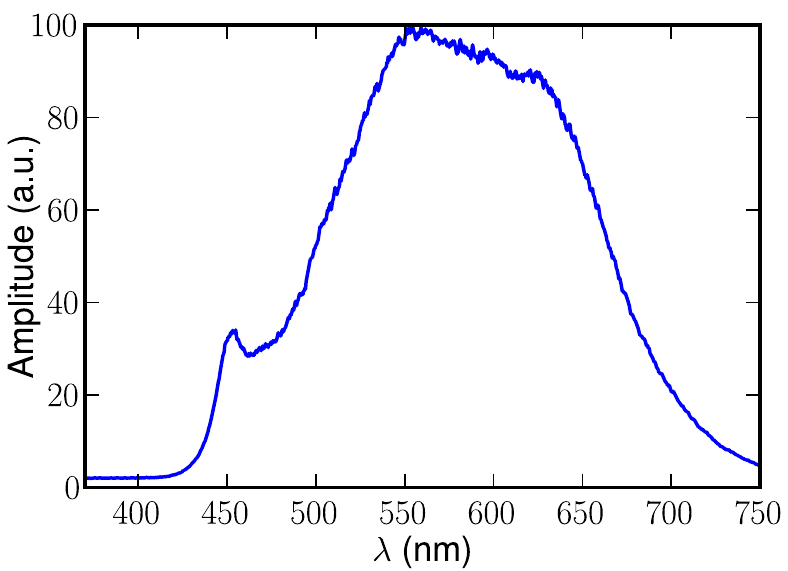} %
  \caption{Measured spectrum of the light source used in our experiments.}
  \label{gr:Spec}
\end{figure}

\subsection*{Supplementary References}
\providecommand*{\mcitethebibliography}{\thebibliography}
\csname @ifundefined\endcsname{endmcitethebibliography}
{\let\endmcitethebibliography\endthebibliography}{}



\end{small}
\end{document}